\documentclass[a4paper,12pt]{article}
\usepackage{longtable,array}
 \usepackage[dvips]{graphicx}
  \usepackage{bm}
   \usepackage{amsmath,amsthm}
    \usepackage{amssymb}
     \usepackage{pifont}
      \usepackage{srcltx}

\newcommand{\nn}{\nonumber}

\newcommand{\tr}{\mathrm{tr}}

\renewcommand{\(}{\left(}
\renewcommand{\)}{\right)}
\renewcommand{\[}{\left[}
\renewcommand{\]}{\right]}

\newcommand{\chil}[1]{\stackrel{o}{#1}}
\newcommand{\chilI}[1]{\stackrel{o}{#1}\raisebox{4pt}{\!$\scriptstyle I$}}
\newcommand{\DiracSlash}[1]{\!\not{\! #1}\,}
\newcommand{\partialboth}{\stackrel{\leftrightarrow}{\partial}}

\begin{document}
\title{On chiral corrections to nucleon GPD}
\author{A.M.Moiseeva\footnote{E-mail: alenasmail@gmail.com }, A.A.Vladimirov\footnote{E-mail: vladimirov.aleksey@gmail.com} \\
Institute f\"ur Theoretische Physik II,\\ Ruhr-Universit\"at Bochum, D-44780, Germany}
\date{}

\maketitle
\begin{abstract}
Within the pion-nucleon chiral perturbation theory we derive the leading chiral correction to the nucleon GPD at $\xi=0$.  We discuss the
difficulties of consideration of nonlocal light-cone operators within the theory with a heavy particle and the methods to solve the
difficulties. The consideration of the chiral corrections directly for nonlocal operators allows to resolve the ambiguity of the inverse Mellin
transformation. In particular, we show that the mixing between  axial and vector GPDs are of order $m_\pi^2/M_N^2$, which is two orders of
magnitude less that it follows from the Mellin moments calculation.
\end{abstract}

\section{Introduction}

In the last decade many papers were devoted to calculations of the low energy properties of generalized parton distributions (GPDs),
\cite{Belitsky:2002jp,Kivel:2004bb,Diehl:2006js,Diehl:2006ya,Ando:2006sk,Dorati:2007bk,Diehl:2005rn,Kivel:2002ia,Kivel:2007jj,Kivel:2008ry}. The
main tool to obtain the low-energy expansion of matrix elements is effective field theories based on spontaneous symmetry breaking, such as
Chiral Perturbation Theory (ChPT) or Heavy Baryon Chiral Perturbation Theory (HBChPT). Investigation of the chiral structure of a hadron is an
important task. Only in this way one can obtain the quark mass dependence and the momentum transfer dependence for GPDs in a model-independent
way. Investigation of the chiral expansion gives us the behavior of a parton distribution at large impact parameters, where it is governed by
pion cloud, see. e.g. \cite{Strikman:2009bd,Alberg:2012wr}. While the pion parton distributions were examined comprehensively, the nucleon GPDs
were considered only as Mellin moments. This article is devoted to the exploration of the chiral corrections to the nucleon GPDs in the
$x$-space ($x$ is the longitudinal hadron momentum fraction carried by a parton).

There is a principal difference between the chiral expansions of a parton distribution $q(x)$ in $x$-space  and its Mellin moments $M_N$. The
point is that the Mellin transformation does not preserve the chiral order of expansion. Let us dwell on it. The expressions for $q(x)$ and
$M_N$ are connected unambiguously via forward and inverse Mellin transformations
\begin{eqnarray}
M_N=\int_0^1 dx ~q(x)~x^N,~~~~~~q(x)=\frac{1}{2\pi i}\int_{c-i\infty}^{c+i\infty}dN~M_N~ x^{-N-1}.
\end{eqnarray}

Using an effective field theory one gets the first terms of the gradient expansion of Mellin moments: $\overline M_N=m^{(0)}_N+a_\chi
m^{(1)}_N$,  where $a_\chi=\frac{m_\pi^2}{\Lambda_\chi^2}$ ($\Lambda_\chi\sim M\sim 4\pi F_\pi\sim 1$GeV is the character scale of the chiral
expansion). The remainder of expansion is of order $a_\chi^2$: $M_N-\overline M_N=\mathcal{O}(a_\chi^2)$. Let us consider the function
$\overline q(x)$, which is a result of  the inverse Mellin transform for $\overline M_N$. The general statement is that the deviation $\delta
q(x)=q(x)-\overline q(x)$ is not necessarily $\mathcal{O}(a_\chi^2)$. The order of the function $\delta q(x)$ depends on $x$ and can not be
guessed without investigation of the analytical property of the full function $q(x)$.

In the case of QCD perturbative expansion, the deviation $\delta q(x)$ is known to be of the  same order as a reminder everywhere except very
small $x$. For the gradient expansion the situation is more sharp, because the variable $x$ is of kinematic origin and expansion over kinematic
parameter can significantly change the analytic properties of a parton distribution. It is clear that the order of deviation $\delta q(x)$
increases with the decreasing of $x$. It follows from the fact that singularities of $q(x)$ are concentrated in the region $x<0$ and for large
$x$ they are shadowed by the singularity at $x=0$. Indeed the analysis of the pion parton distribution in the ChPT
\cite{Kivel:2007jj,Kivel:2008ry} shows that the deviation $\delta q(x)$ is $\mathcal{O}(a_\chi^2)$ for $x\gtrsim a_\chi$, but became
$\mathcal{O}(a_\chi)$ at $x\lesssim a_\chi$. In order to reach the declared accuracy $\mathcal{O}(a_\chi^2)$ at $x\lesssim a_\chi$, the
contribution of particular class of diagrams should be added. The calculation of the low-$x$ contribution to the pion GPD was performed with the
help of recursive equation technic \cite{Kivel:2008mf}, and leads to evident effects \cite{Perevalova:2011qi}. For the nucleon target such an
analysis was never done.

The starting point of the analysis is the operator definition of GPD. GPDs are defined as matrix elements of the light-cone QCD operator.  The
light-cone separation $\lambda$ is dimensional, and hence has its own counting rule in the chiral expansion. Contribution of some particular
region of $\lambda$-integration is distinct for different $x$. This allows one to correlate $\lambda$-counting rules with an appropriate
$x$-region of GPD. In the Mellin moments one integrates over all range of $x$ and therefore mixes up distinct $\lambda$-counting rules. In order
to resolve this ambiguity, it is convenient to transfer the counting   prescription of $\lambda$ onto some other object, which is not washed out
during Mellin transformation. The natural object is the vector of light cone direction $n_\mu$ ($n^2=1$): first of all, it accompanies the
parameter $\lambda$ in the GPD operator, second, the normalization of the vector $n$ can be changed without changing of GPD definition:
\begin{eqnarray}\label{norm_n}
n_\mu\rightarrow \Lambda n_\mu,~~~~~\bar n_\mu\rightarrow \frac{1}{\Lambda}\bar n_\mu,~~~~n_\mu \bar n^\mu=1.
\end{eqnarray}
Therefore, adjusting the special counting rules for the vector $n$($\bar n$) one obtains the expression for $\overline M_N$ such that $\delta
q(x)=\mathcal{O}(a_\chi^2)$ in the corresponded area of $x$. The detailed analysis of the counting rules for vector $n$ in different regions of
$x$ is presented in the sec.II.

Notice, that applying the special counting rules for vector $n$ the  contributions of the operators with higher number of derivatives to Mellin
moments can be allowed, contrary the standard counting rules. Then the expression for $\overline M_N$ is not truncated by $a_\chi$ term, but can
contain terms with higher orders of $a_\chi$. This exceedance of the accuracy is needed for the restoration of the proper analytic properties of
the distribution. It also shows that the calculation directly with nonlocal operators looks preferably. In the later case one should not care on
the special counting rules because the parameter $\lambda$ is not integrated out, and its size can be directly taken into account. The price is
the incorporation of the nonlocal operator in the gradient expansion.

Additional difficulty arises from to the power counting problem of the pion-nucleon ChPT. The standard tool to avoid the counting problem is to
use HBChPT, which is a non-relativistic limit of ChPT. However, naive attempt to consider a light-cone operator in the non-relativistic
framework is condemned to fail. The infrared logarithms of straightforward HBChPT are replaced by some cumbersome functions, which often have
artificial singularities in $x$. The reason is the non-commutativity of the non-relativistic limit and the infinite momentum frame nature of the
light-cone operator. Therefore, one should consider the nonlocal operator in the covariant framework with consequent procedure of the
redefinition of the higher terms by large-mass corrections \cite{Schindler:2003xv,Fuchs:2003qc}. We notice that the heavy baryon reduction via
the infrared regulation \cite{Becher:1999he}, does not suite here as well, because the loop-integrals with nonlocal operator have infrared
properties distinct from the properties of  usual loop-integral. We describe the non-contradictory heavy baryon reduction for loop-integral with
nonlocal operators in sec. III.

In this paper we calculate the leading non-analytical chiral correction to nucleon GPDs at $\xi=0$ directly from nonlocal operators in the
relativistic covariant framework using extended on-mass-shell subtruction scheme, EOMS \cite{Schindler:2003xv,Fuchs:2003qc}. The structure of
the article is following: in sec.II we derive the effective nonlocal operators and show the difference between our operator and the operators
constructed in HBChPT. In sec. III we perform the detailed investigation of the heavy baryon limit for the nonlocal operators, and derive rules
for self-consistent nonlocal operator consideration within ChPT. The results of our calculations are presented in sec.IV.

\section{Nucleon GPD in the chiral perturbation theory}

Nucleon GPDs parameterize the matrix elements of nonlocal operators:
\begin{eqnarray}\label{V_QCD}
&&\int\frac{d\lambda}{2\pi}e^{-ix\lambda P_+}
\langle p'|\bar q(\frac{\lambda n}{2})\DiracSlash{n} \tau^Aq(-\frac{\lambda n}{2})|p\rangle~=~
\\ \nn &&~~~~~~~~~~~~~~~
\frac{1}{P_+} \bar u(p')\Bigg[\DiracSlash n H(x,\xi,\Delta^2)+\frac{i\sigma^{\mu\nu}n_\mu\Delta_\nu}{2M}E(x,\xi,\Delta^2)\Bigg]\tau^Au(p),
\\\label{A_QCD}
&&\int\frac{d\lambda}{2\pi}e^{-ix\lambda P_+}
\langle p'|\bar q(\frac{\lambda n}{2})\DiracSlash n \gamma_5\tau^Aq(-\frac{\lambda n}{2})|p\rangle~=~
\\ \nn &&
~~~~~~~~~~~~~~~
\frac{1}{P_+}\bar u(p')\Bigg[\DiracSlash n \gamma_5 \widetilde H(x,\xi,\Delta^2)+\gamma_5\frac{(n\Delta)}{2M}\widetilde E(x,\xi,\Delta^2)\Bigg]\tau^Au(p),
\end{eqnarray}
where $n_\mu$ is a light-cone vector projecting the ``plus'' component of momenta; $\bar u$ and $u$ are nucleon spinors; $M$ is a nucleon mass.
We use the standard notation for kinematical variables in the Breit reference frame:
$$
p'=P+\frac{\Delta}{2},~~~p=P-\frac{\Delta}{2},~~~\xi=-\frac{\Delta_+}{2P_+}.
$$
The variable $x$ has the meaning of a momentum fraction carried by a
quark with respect to the average momentum $P$ of the nucleon. The
index $A$ is equal to zero for isospin scalar GPDs, and $A=1,2,3$
for isospin vector GPDs.

ChPT is the low-energy effective field theory of QCD. The lowest-order pion-nucleon Lagrangian reads:
\begin{eqnarray}\label{complete_L}
\mathcal{L}=\bar \Psi\(i\gamma^\mu(\partial_\mu+\Gamma_\mu)-M+\frac{g_a}{2}u_\mu\gamma^\mu\gamma^5\)\Psi+
\frac{F^2_\pi}{4}\tr\[\partial_\mu U \partial^\mu U^\dagger+\chi^\dagger U+\chi U^\dagger)\]
\end{eqnarray}
with $M$ being the nucleon mass and $F_\pi\simeq 93$ MeV being the
pion decay constant and $\chi$ being the quark mass matrix. We use
the standard notation for the pion field constructions
$$u^2=U=e^{\frac{i\pi^a\tau^a}{F_\pi}},~~~~\Gamma_\mu=\frac{1}{2}\[u^\dagger,\partial_\mu u\],~~~~u_\mu=i u^\dagger \partial_\mu U u^\dagger.$$

Calculations with Lagrangian (\ref{complete_L}) lead to the power-counting problem, since the nucleon mass is of the same order as the normalization scale of ChPT, $M\sim \Lambda_{\chi}\sim 1$ GeV. The most popular solution of the counting problem is provided by HBChPT \cite{1}. The nucleon field is split onto soft and hard components:
\begin{eqnarray}\label{field_rescaling}
\Psi(x)=e^{-iM(vx)}\(\mathcal{N}_v+\mathcal{H}_v\),~~~~\mathcal{N}_v(x)=e^{i M(vx)}\frac{1+\DiracSlash v}{2}\Psi(x)
\end{eqnarray}
where $v_\mu$ is the nucleon velocity vector, $v^2=1$. One eliminates the component $\mathcal{H}_v$ using the equations of motion and obtains HBChPT Lagrangian. The pion-nucleon part of HBChPT Lagrangian reads
\begin{eqnarray}\label{S_HBChPT}
\mathcal{L}_{N\pi}=\bar{\mathcal{N}}_v\(i v^\mu (\partial_\mu+\Gamma_\mu)+g_a  S^\mu u_\mu\)\mathcal{N}_v,
\end{eqnarray}
where $S_\mu=\frac{i}{2}\gamma^5\sigma^{\mu\nu}v_\nu$ is the spin operator, $(vS)=0$.

Right-hand-side of the matrix elements (\ref{V_QCD}-\ref{A_QCD}) in the heavy baryon limit becomes:
\begin{eqnarray}\label{V_HQCD}
&&\int\frac{d\lambda}{2\pi}e^{-ix\lambda P_+}
\langle p'|\bar q(\frac{\lambda n}{2})\DiracSlash n \tau^Aq(-\frac{\lambda n}{2})|p\rangle~=~
\\ \nn
&&~~~~~~~~~~~~~~~~~~~~~~~~
\frac{1}{P_+}\[v_+\bar N_v \tau^A N_v\(H+\frac{\Delta^2}{4M^2}E\)+\frac{H+E}{M}\bar N_v \tau^A[S_+,(S\Delta)] N_v\]
\\\label{A_HQCD}
&& \int\frac{d\lambda}{2\pi}e^{-ix\lambda P_+}
\langle p'|\bar q(\frac{\lambda n}{2})\DiracSlash n \gamma^5\tau^Aq(-\frac{\lambda n}{2})|p\rangle~=~
\\ \nn
&&~~~~~~~~~~~~~~~~~~~~~~~~
\frac{1}{P_+}\[ 2\gamma\widetilde H~\bar N_v S_+ \tau^AN_v+\(\widetilde E+\frac{\widetilde H}{1+\gamma}\)\bar N_v \frac{\tau^A \Delta_+(S\Delta)}{2M^2}N_v\],
\end{eqnarray}
where $N_v$ is the heavy component of the nucleon spinor: $N_v(p)=\sqrt{\frac{2}{1+\gamma}}\frac{1+\not v}{2}u(p)$, $\gamma=\sqrt{1+\frac{\Delta^2}{4M^2}}$.

In the limit $M\gg m$, the parameter $\xi\sim\frac{m}{M}\ll 1$. In this article we put $\xi$ to zero, by setting $\Delta_+=0$. The effects related to non-zero $\xi$ can be of the leading order in $M^{-1}$ expansion, and they will be considered in a separate article. In the limit $M\rightarrow \infty$ the GPDs are
\begin{eqnarray}
H(x,0,\Delta^2)=q(x,\Delta^2),~~~~~\widetilde H(x,0,\Delta^2)=\Delta q(x,\Delta^2),~~~~~E(x,0,\Delta^2)=E(x,\Delta^2),
\end{eqnarray}
where $q$ and $\Delta q$ are the parton distribution functions (PDFs) with transverse momentum dependance. GPD $\widetilde E$  can not be obtain in the limit $\Delta_+=0$.

\subsection{Matching of nonlocal operators}

In this subsection we construct the effective vector and axial
light-cone operators for the nucleon. Our procedure is very close to
one presented in \cite{Kivel:2004bb,Kivel:2002ia} with minor
redefinitions.

The procedure of the matching QCD light-cone operators to the operators in the chiral effective theories is well-known. One should introduce the
nonlocal operator with light-cone separation $\lambda$ and transformation properties of the initial operator. We start with QCD operators:
\begin{eqnarray}\label{O}
O^A(\lambda) &=&\bar q(\frac{\lambda n}{2})\DiracSlash n \tau^Aq(-\frac{\lambda n}{2}),
\\ \label{O5}
\widetilde{O}^A(\lambda) &=&\bar q(\frac{\lambda n}{2})\DiracSlash n \gamma_5 \tau^Aq(-\frac{\lambda n}{2}).
\end{eqnarray}
These operators transform in mixed representation of the chiral
rotations group. It is convenient to deal with the left- and
right-handed light-cone operators:
$$
O_R^A(\lambda)=O^A(\lambda)+\widetilde{O}^A(\lambda),~~~~
O_L^A(\lambda)=O^A(\lambda)-\widetilde{O}^A(\lambda),
$$
which transform as the define representation.

The fields involved in the chiral Lagrangian transforms under local $SU_L(2)\times SU_R(2)$ as:
\begin{eqnarray}
U\rightarrow R~ U~ L^\dagger,& & \chi \rightarrow R~\chi~ L^\dagger, \\
\Psi\rightarrow K(u,R,L)~\Psi, & ~~~~~~~~~& u\rightarrow R~ u~K^\dagger(u,R,L)=K(u,R,L)~ u~ L^\dagger,
\end{eqnarray}
where $R$ and $L$ are  matrices of right and left chiral rotations, and $K$ is a matrix of chiral transformations for nucleon field containing field $u$.

The lowest order operators contain no derivatives. The most general structures respecting the symmetries are:
\begin{eqnarray*}
&&O^A_R(\lambda)~=~\\
&&\int d\beta d\alpha \(F^I_1(\beta,\alpha) ~\bar\Psi(x) u^\dagger(x)\DiracSlash n
\frac{\tau^A}{2} u(y)\Psi(y)+F^I_2(\beta,\alpha)~\bar\Psi(x) u^\dagger(x)\DiracSlash n
\gamma^5\frac{\tau^A}{2} u(y)\Psi(y)\),
\\
&&O_L^A(\lambda)~=~ \\
&& \int d\beta d\alpha \(F^I_3(\beta,\alpha) ~\bar\Psi(x) u(x)\DiracSlash n
\frac{\tau^A}{2} u^\dagger(y)\Psi(y)+F^I_4(\beta,\alpha)~\bar\Psi(x) u(x)\DiracSlash n
\gamma^5\frac{\tau^A}{2} u^\dagger(y)\Psi(y)\),
\end{eqnarray*}
where $x=\lambda n \frac{\alpha+\beta}{2}$ and $y=\lambda n
\frac{\alpha-\beta}{2}$. The integration area over $\alpha$ and
$\beta$ is $|\alpha|+\beta<1$, these limits are provided by
additional requirement of polynomiality. The functions
$F^I(\beta,\alpha)$ represent generation functions for the tower of
low energy constants, and have the meaning of the double
distributions for the nucleon GPDs with given isospin number in the
chiral limit \cite{Radyushkin:1998rt}.

The properties of operators (\ref{O},\ref{O5}) under parity
transformations demand that $F^I_1=F^I_3$ and $F^I_2=-F^I_4$. Thus,
the expression for vector and axial light-cone operators are
\begin{eqnarray}\label{O_in_ChPT}
O^A_{N\pi}(\lambda)&=&\int d\beta d\alpha
\,\frac{1}{2}\(F^I_1(\beta,\alpha) \bar\Psi(x) \DiracSlash{n} t_+^A(x,y)\Psi(y)+F^I_2(\beta,\alpha) \bar\Psi(x) \DiracSlash n\gamma^5 t_-^A(x,y)\Psi(y)\), \\
\label{A_in_ChPT}
\widetilde O^A_{N\pi}(\lambda)&=&
\int d\beta d\alpha
\,\frac{1}{2}\(F^I_2(\beta,\alpha) \bar\Psi(x) \DiracSlash n\gamma^5 t_+^A(x,y)\Psi(y)+F^I_1(\beta,\alpha) \bar\Psi(x) \DiracSlash nt_-^A(x,y)\Psi(y)\),
\end{eqnarray}
where the scalar and pseudo-scalar combinations of pion fields are
$$t^A_\pm(x,y)=u^\dagger(x)\tau^A u(y)\pm u(x)\tau^A u^\dagger(y).$$

In expressions (\ref{O_in_ChPT},\ref{A_in_ChPT}) one has freedom to
choose the normalization of the generating function in a convenient
way. To recover that the $F_{1,2}$ are the double distributions in
the chiral order one should calculate the matrix element $\int
\frac{d\lambda}{2\pi}e^{-ix\lambda P_+}\langle
p'|O_{N\pi}^A|p\rangle$ at the tree level, and compare the result
with definition (\ref{V_QCD}). We obtain:
\begin{eqnarray}\label{F_Interpretation0}
\int_{-1}^1 d\beta \int_{-1+|\beta|}^{1-|\beta|}d\alpha
~F^I_1(\beta,\alpha)\delta(x-\alpha\xi-\beta)=\chil H^I(x,\xi),
\end{eqnarray}
where the symbol $\chil {~}$ denotes the chiral limit of a quantity. The similar consideration for function $F_2(\beta,\alpha)$ gives:
\begin{eqnarray}
\int_{-1}^1 d\beta \int_{-1+|\beta|}^{1-|\beta|}d\alpha
~F^I_2(\beta,\alpha)\delta(x-\alpha\xi-\beta)=\chilI{\widetilde
H}(x,\xi).
\end{eqnarray}
In the limit $\xi=0$, which is of the special interest in this paper, the $\alpha$-dependence of the operators can be moved out by the translation. The resulting generating functions are
\begin{eqnarray}\label{F-interpretation}
\int_{-1+|\beta|}^{1-|\beta|}d\alpha F^I_1(\beta,\alpha)=\chilI{q}(\beta),~~~~
\int_{-1+|\beta|}^{1-|\beta|}d\alpha F^I_2(\beta,\alpha)=\chilI{\Delta q}(\beta),
\end{eqnarray}
where $\chil{q}(\beta)$ and $\chil{\Delta q}(\beta)$  are the nucleon PDFs in the chiral limit. The normalization of the PDFs follows from the vector and axial current operators, it reads:
\begin{eqnarray}
\label{F-normalization}
\int_{-1}^{1}d\beta \chilI{q}(\beta)=1,~~~~~~~
\int_{-1}^{1}d\beta \chil{\Delta q}(\beta)=g_a.
\end{eqnarray}

The reduction of the operators (\ref{O_in_ChPT},\ref{A_in_ChPT}) to the heavy-baryon form is straightforward:
\begin{eqnarray}\label{O_in_HBChPT}
O^A_{N\pi}(\lambda)&=&\int d\beta d\alpha e^{iM(v(x-y))}\Bigg(\frac{v_+}{2}F^I_1(\beta,\alpha) ~\bar{\mathcal{N}}_v(x)t_+^A(x,y)\mathcal{N}_v(y)
\\ \nn&&~~~~~~~~~~~~~~~~~~~~~+
F^I_2(\beta,\alpha) ~\bar{\mathcal{N}}_v(x) S_+ t_-^A(x,y)\mathcal{N}_v(y)\Bigg)+\mathcal{O}\(\frac{1}{M}\),
\end{eqnarray}
\begin{eqnarray}\label{A_in_HBChPT}
\tilde O^A_{N\pi}(\lambda)&=&\int d\beta d\alpha e^{iM(v(x-y))} \Bigg(\frac{v_+}{2}F^I_1(\beta,\alpha)  ~\bar{\mathcal{N}}_v(x) t_-^A(x,y)\mathcal{N}_v(y)
\\ \nn&&~~~~~~~~~~~~~~~~~~~~+
F^I_2(\beta,\alpha)~ \bar{\mathcal{N}}_v(x) S_+ t_+^A(x,y) \mathcal{N}_v(y)\Bigg)+\mathcal{O}\(\frac{1}{M}\).
\end{eqnarray}

We remind that in HBChPT nucleons are taken to be non-relativistic, and nucleon momenta which flow through the diagrams are reduced momenta: $r_\mu=p_\mu-Mv_\mu$. The expressions for operators (\ref{O_in_HBChPT},\ref{A_in_HBChPT}) have the exponential factor $\exp(iM(v(y-x)))$, which restores the reduced nucleon momentum to the complete nucleon momentum in the operator vertex. In principal, this factor can be absorbed to the definition of the low-energy generating functions $F_{1,2}$, but this would lead to the inconvenient scale of parameters $\beta$ and $\alpha$. We keep the exponential factors as they are, in order to have the simple interpretation (\ref{F_Interpretation0}-\ref{F-normalization}) for the generating functions.

The matrix element of operators (\ref{O_in_ChPT}-\ref{A_in_ChPT}) contributes to the GPDs $H$ and $\widetilde H$ at the tree order, and to all set of GPDs at one-loop order. The leading contribution to the GPDs $E$ and $\widetilde E$ is given by the operators of the higher order in chiral counting. The operator giving a chiral limit of GPD $E$ is proportional to $n_\mu\partial_\nu(\bar \Psi \sigma^{\mu\nu}\Psi)$, and it does not give a contribution to $H$ and $\widetilde H$ at tree and one-loop level. Therefore, in this article, which is mostly devoted to GPDs $H$ and $\widetilde H$ we leave such kind of operators out.

Pure pion nonlocal operators respecting the transformation properties of (\ref{O},\ref{O5}) can be constructed. Their form and properties were
derived at \cite{Kivel:2002ia}. The pure pion operators for the vector and axial PDFs have the form:
\begin{eqnarray}\label{pion-OP}
O^A_{\pi\pi}(\lambda)&=&\frac{-iF_\pi^2}{4} \int d\beta d\alpha ~\mathcal{F}(\beta,\alpha) \tr\[\tau^A\(U(x)\overleftrightarrow{\partial}_+U^\dagger(y)+U^\dagger(x)\overleftrightarrow{\partial}_+U(y)\)\],
\end{eqnarray}
\begin{eqnarray}\label{pion-O5}
\widetilde O^A_{\pi\pi}(\lambda)&=&\frac{-iF_\pi^2}{4} \int d\beta d\alpha ~\widetilde{\mathcal{F}}(\beta,\alpha) \tr\[\tau^A\(U(x)\overleftrightarrow{\partial}_+U^\dagger(y)-U^\dagger(x)\overleftrightarrow{\partial}_+U(y)\)\],
\end{eqnarray}
where the integration area for $\alpha$ and $\beta$ is $|\alpha|+\beta<1$, and $\overleftrightarrow{\partial}=\overrightarrow{\partial}-\overleftarrow{\partial}$. The axial pion operator gives no contribution at $\Delta_+=0$ (up to one-loop level), and therefore, we will skip it in current consideration.

The low-energy generation function $\mathcal{F}(\beta,\alpha)$ belongs to the pion GPD in the chiral limit via the expressions:
\begin{eqnarray}
\chil{H^I}(x,\xi)=\int_{-1}^1 d\beta \int_{-1+\beta}^{1-\beta}d\alpha d\beta \mathcal{F}^I(\beta,\alpha)
\[\delta(x-\alpha\xi-\beta)-(1-I)\xi \delta\(x-\xi(\alpha+\beta)\)\],
\end{eqnarray}
where $I$ stands for the isospin, and $\mathcal{F}^{I=1(0)}(\beta,\alpha)=\frac{1}{2}\(\mathcal{F}(\beta,\alpha)+(-)\mathcal{F}(-\beta,\alpha)\)$.
For the forward limit of the pion PDF  we will use the notation $\chil{H^I}(x,\xi=0)=\chilI Q(x)$. The normalization of pion PDFs is:
\begin{eqnarray}
\int_{-1}^1 d\beta \stackrel{o}{Q}\raisebox{6pt}{\!$\scriptstyle I$}(\beta)~=~\delta^{I,1}.
\end{eqnarray}

\subsection{Operators for the Mellin moments}

The Mellin moments of GPDs are widely used in phenomenology. The operator of the Mellin moment can be found using the well-known relation:
\begin{eqnarray}\label{Op<>M_QCD}
\int_{-1}^1 dx ~ x^N \int \frac{d\lambda}{2\pi}e^{-i\lambda x P_+}~ \bar q\(-\frac{\lambda n}{2}\)\gamma^+ q \(\frac{\lambda n}{2}\) =
\frac{1}{P_+^{N+1}}\bar q(0)\gamma^+\(-i \partialboth_+\)^Nq(0).
\end{eqnarray}
Calculation of the Mellin moments in effective field theory is straightforward and was performed in \cite{Diehl:2006js,Diehl:2006ya,Ando:2006sk,Dorati:2007bk} up to second chiral order. The result of calculation has the form:
\begin{eqnarray}\label{M_form}
\overline M_N=m^{(0)}_N+a_\chi m^{(1)}_N,
\end{eqnarray}
where $a_\chi$ is the chiral expansion parameter, $a_\chi\sim \frac{m^2}{(4\pi F_\pi)^2}$. The reminder  term of the expansion is $M_N-\overline
M_N=\mathcal{O}(a_\chi^2)$.  However, the function $\delta q(x)=q(x)-\overline q(x)$ (here, $q(x)$ is the full parton distribution and
$\overline q(x)$ is the reconstructed from $\overline M_N$ parton distribution)  is not necessarily $\mathcal{O}(a_\chi^2)$ in the full domain
of $x$. The main reason for non-commutation of the perturbative expansion and the Mellin transformation is the violation of the analytical
properties of $q(x)$ by the chiral expansion. There  are no doubts, that the low-energy expansion changes the analytical structure of the parton
distributions, and the inverse Mellin transformation of the low-energy expansion has different significance in different regions of $x$.

The left-hand-side of equation (\ref{Op<>M_QCD}) contains the dimensional parameter $\lambda$, which has  its own chiral counting. The different
areas of $\lambda$ are responsible for the physical content of the different regions of $x$.  Performing Mellin transformation one integrates
over all range of $x$ and therefore mixes up various counting rules for $\lambda$. In order to resolve this ambiguity, it is convenient to refer
the counting  prescription of $\lambda$ to the light-cone vector $n_\mu$. This can be done via changing the normalization factor $\Lambda$ of
the vector $n$ in eqn.(\ref{norm_n}). Adjusting special counting rule to the vector $n$ we obtain the expression for $\overline M_N$ such that
$\delta q(x)=\mathcal{O}(a_\chi^2)$ in the corresponding area of $x$. General form of $\overline q(x)$ reads:
\begin{eqnarray}\label{q_form}
\overline q(x)=\chil q(x)+a_\chi \int_{-1}^1 \frac{d\beta}{\beta} \chil q\(\frac{x}{\beta}\) C(\beta,a_\chi),
\end{eqnarray}
where the function $C(\beta,a_\chi)$ can not be expanded in $a_\chi$  without distraction of the analytical structure. As it will be
demonstrated later, often the Mellin convolution in (\ref{q_form}) is well-behaved in total, but it is divergent for every term of Taylor
expansion of $C(\beta,a_\chi)$ at $a_\chi=0$. Performing Mellin transformation of (\ref{q_form}) one gets:
\begin{eqnarray}\label{demonstration of m}
\int dx~ x^{N} ~\overline q(x)=m^{(0)}_N+a_\chi \tilde m_N(a_\chi),
\end{eqnarray}
where $\tilde m_N(a_\chi)$ can be further expanded over $a_\chi$.  The term $\tilde m_N(0)$ is equal to $m^{(1)}_N$ and, therefore, the
expression (\ref{demonstration of m}) coincides with the expression (\ref{M_form}) at order $\mathcal{O}(a_\chi^2)$. The exceedance of the
accuracy in (\ref{demonstration of m}) is needed for the restoration of the proper analytic properties of the distribution $\overline q(x)$.

We start the consideration of the counting rules for $\lambda$ from the analysis of the standard counting rules of HBChPT, where the light-cone
vector $n_\mu$ has no special prescription. The standard counting rules of HBChPT reads:
\begin{eqnarray}\label{standard_counting_rules}
v_\mu \sim \mathcal{O}(1),&&\partial_\mu \pi \sim \mathcal{O}(q),~~ \partial_\mu\mathcal{N}_v \sim \mathcal{O}(q)
\end{eqnarray}
where $q$ is the parameter of gradient expansion, i.e. small momenta or the pion mass, $q\sim m_\pi$. In the HBChPT nonlocal nucleon operators
also contain the exponent with the heavy nucleon mass, which suppresses soft derivatives. Therefore, in the leading order the HBChPT operator
(\ref{O_in_HBChPT}) reads:
\begin{eqnarray}
&&O^A_{\text{rules (\ref{standard_counting_rules})}}(\lambda)=
\\ &&\nn
\int d\beta d\alpha~ e^{i\lambda M v_+\beta}\Bigg(v_+F_1(\beta,\alpha) ~\bar{\mathcal{N}}_v(0)t_+^A(0)\mathcal{N}_v(0)+
2F_2(\beta,\alpha)~\bar{\mathcal{N}}_v(0)S_+ t_-^A(0)\mathcal{N}_v(0)\Bigg).
\end{eqnarray}
Taking into account that $Mv_+=P_+$ and (\ref{F-interpretation}) we found
\begin{eqnarray}\label{O_in_(25)}
P_+\int \frac{d\lambda}{2\pi}e^{-i x P_+\lambda}O^A_{\text{rules (\ref{standard_counting_rules})}}(\lambda)=
v_+\chil q(x)~\bar{\mathcal{N}}_v(0)t_+^A(0)\mathcal{N}_v(0)+
2\chil{\Delta q}(x)~ \bar{\mathcal{N}}_v(0)S_+ t_-^A(0)\mathcal{N}_v(0).
\end{eqnarray}
Operator (\ref{O_in_(25)}) allows one to obtain only such corrections to PDF
 those do not touch the intrinsic dynamics of the nucleon. In other words using rules (\ref{standard_counting_rules}) one indirectly uses the approximation $q(x,\Delta^2)=q(x)F(\Delta^2)$, where $F(\Delta^2)$ is the form-factor with corresponding quantum numbers. The standard counting rule $n\sim 1$ relates to the region of $x$ close to unity. That is, at $x\sim 1$ the reconstructed parton distribution $\overline q(x)$ deviates as $\mathcal{O}(a_\chi^2)$ from the full one. In this region the short $\lambda$ ($\lambda\sim \frac{1}{xP_+}\sim\frac{1}{M}$) dominates in the integral (\ref{Op<>M_QCD}). 


In the region $x\sim \frac{q}{M}$ we need to set the condition $n\sim q^{-1}$ (this follows from the condition that the argument of the exponent should be of order unity, otherwise integral is suppressed). As a result one has the following counting rules:
\begin{eqnarray}\label{new_counting_rules1}
\partial_+\pi \sim \mathcal{O}\(1\),~~\partial_-\pi \sim \mathcal{O}\(q^2\),~~\partial_\perp \pi \sim \mathcal{O}\(q\),
\end{eqnarray}
and similar for the nucleon field. Also the consistent counting rules should be attached to the vector $v_\mu$
\begin{eqnarray}\label{new_counting_rules2}
v_+\sim \mathcal{O}\(\frac{1}{M}\).
\end{eqnarray}
In this case all light-cone derivatives are of the same importance. The operators for Mellin moments can be obtained directly by the expansion
of the operators (\ref{O_in_HBChPT},\ref{A_in_HBChPT}). Numerically this area corresponds to $x\sim\frac{m_\pi}{M}\sim0.15$. In this region the
loops with composite operators should be calculated with a great care, because usually ultraviolet logarithms of $xP_+$ can become infrared.
And, indeed, such effects take a place and we discuss them in the next section.

The region $x\sim \frac{q}{M^2}$ is governed by the higher derivative  terms since $n\sim \frac{M}{q}$ and $\partial_+\pi\sim M$. The integral
(\ref{Op<>M_QCD}) is controlled by the huge light-cone distance $\lambda\sim \frac{M}{q^2}$. Therefore, the higher terms of the chiral expansion
are of the same importance as the leading order and should be taken into account simultaneously. The terms belonging to this region have the
form of $\delta$-functions and their derivatives, see the case of pion \cite{Kivel:2007jj,Kivel:2008ry}. The resummation of such series in all
orders of chiral expansion leads to the leading term, which describes the long-range behavior of the pion cloud. This mechanism is also known as
the ``chiral inflation'', see \cite{Perevalova:2011qi}.

We stress, that many problems with intermediate regions can be avoided performing calculation directly with nonlocal operators. Keeping the
parameter $\lambda$ alive one leaves the decision of the interesting $x$-region until the final result. As a price one has to perform more
involved calculation.

\section{Nonlocal operators in the heavy baryon approach}

\begin{figure}[t]
\begin{center}
\includegraphics[width=0.7\textwidth]{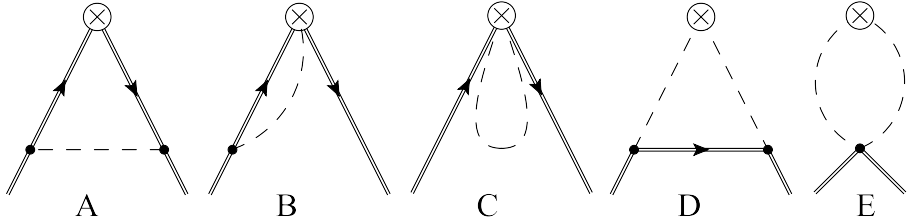}
\end{center}
\caption{Five diagrams contributing to the leading order. The solid and the dashed lines denote proton and pion fields, respectively.}
\end{figure}

There are five diagrams contributing to the leading order of PDF chiral expansion, fig.1.  The calculation of these diagrams is almost
straightforward. However, the final result contains some uncertainties, related to non-commutativity of the heavy baryon limit and light-cone
definition of the operator, which we discuss in this section.

To show features of the calculation with nonlocal operators in heavy baryon theories we  consider in details the diagram B. For concreteness, we
consider the vector isovector operator. The diagram B is interesting because it mixes the axial and the vector PDFs with each other. Such a
mixing is believed to be strongly suppressed, because the Mellin moments of this diagram are zero at leading order, due to convolution of the
nucleon spin-vector with the nucleon velocity. Whereas the calculation with nonlocal operator gives non-zero result already at the leading
order.

The straightforward calculation of loop integral with the Lorentz invariant action (\ref{complete_L}) before the renormalization procedure reads:
\begin{eqnarray}\label{B1}
&&B(x)=
 \frac{g_a}{4}\frac{\bar N\tau^a\hat n N}{P_+}\Gamma(-1+\epsilon)
 \\\nn&&
\times \int_{-1}^1 d\beta\chil{\Delta q}(\beta)
\[\delta(x-\beta)
 \frac{(m^2)^{1-\epsilon}}{(4\pi F_\pi)^2}+\frac{2M^2}{(4\pi F_\pi)^2}\int_0^1 d \eta~\bar \eta \[M^2\bar \eta^2+m^2\eta\]^{-\epsilon} \delta(x-\eta\beta)\],
\end{eqnarray}
where $\bar \eta=1-\eta$, and $\epsilon$ is a parameter of dimensional regularization: $d=4-2\epsilon$. The integral in the brackets gives:
\begin{eqnarray}\label{B11}
2M^2\int_0^1 d \eta~\bar \eta \[M^2\bar \eta^2+m^2\eta\]^{-\epsilon} \delta(x-\eta\beta)
=2(M^2)^{1-\epsilon}~\frac{\bar y}{\beta} \[\bar y^2+\alpha^2y\]^{-\epsilon}\theta(0<y<1),
\end{eqnarray}
where $y=\frac{x}{\beta}$ and $\alpha=\frac{m}{M}$. This part is proportional to the large mass and cannot be considered as a correction without additional adaptation.

The standard approach of the heavy baryon theory posses the rescaling of the nucleon fields (\ref{field_rescaling}) and afterwards expansion in the small ratio $\alpha\sim 0.15$. In the language of the Feynman diagrams it means a partial resummation of the perturbative expansion and it was never done explicitly. The bypass way was suggested in \cite{Becher:1999he}: the heavy baryon result is provided by the infrared singular part of the loop-integral, whereas the infrared regular part is to be subtracted into the higher order coupling constants. This method works perfectly for usual loop-integrals, but faces problems with loop-integrals of type (\ref{B1}). Applying formally the method of \cite{Becher:1999he} to the integral (\ref{B1}) one gets:
\begin{eqnarray}\label{B[17]}
&&B_{\text{\cite{Becher:1999he}}}(x)=
 \frac{g_a}{4}\frac{\bar N\tau^a\hat n N}{P_+}\Gamma(-1+\epsilon)
 \\\nn&&
\times \int_{-1}^1 d\beta\chil{\Delta q}(\beta)
\[\delta(x-\beta)
 \frac{(m^2)^{1-\epsilon}}{(4\pi F_\pi)^2}+\frac{2(M^2)^{1-\epsilon}}{(4\pi F_\pi)^2}\frac{\bar y}{|\beta|}(\bar y^2+\alpha^2 y)^{-\epsilon}\theta(y<1)\].
\end{eqnarray}
This expression is still proportional to $M^2$. The reason is that the $\delta$-function of the nonlocal operator changes the low-energy
behavior of the loop-integral and does not allow to reach the infrared pole at $\alpha\rightarrow0$. Also expression (\ref{B[17]}) contains
incorrect $\theta$-function, which results to the singular integral over $\beta$. The integration over $x$ for expression (\ref{B[17]}) is
infrared divergent, due to the incorrect $\theta$-function. This reproduces the infrared singular part for the loop-integrals for the Mellin
moments.

The proper way to reduce the integral (\ref{B11}) to the heavy baryon limit is not to extract the infrared singular part of the loop-integral,
but to subtract its infrared regular part. The regular part of the loop integral is not changed by the nonlocality, because at the limit
$M\rightarrow \infty$ the delta-function can be moved out of the loop-integral. Therefore, the reduction to the heavy baryon limit for nonlocal
operators consists in the subtraction of the pole at $M\rightarrow \infty$, with the subsequent  renormalization procedure. The described
procedure is known as extended on-mass-shell renormalization \cite{Schindler:2003xv,Fuchs:2003qc}. After the renormalization procedure for the
expression (\ref{B1}) reads:
\begin{eqnarray}\label{B5}
&&B(x)=\frac{\bar N\tau^a\hat n N}{P_+}\int_{-1}^1 \frac{d\beta}{\beta}\chil{\Delta q}(\beta)
\\\nn&&~~~~~~~~~~~~~\times
 \[\frac{g_a}{4}\delta(1-y)\frac{m^2}{(4\pi F_\pi)^2}\ln(m^2)+\frac{g_a}{2}\frac{M^2}{{(4\pi F_\pi)^2}}~\bar y
 \ln\(1+\frac{y\alpha^2}{\bar y^2}\)\theta(0<y<1)\].
\end{eqnarray}
This expression is $\mathcal{O}(\alpha^2)$, in spite of the large mass $M$ in the second term.

The expression (\ref{B5}) has a general form (\ref{q_form}). Expanding the logarithm in the second term in powers of $\alpha$ one get the asymptotic series, because every term starting from $\alpha^2$ is divergent at $\beta=x$ integral due to the factor $\bar y^{-k}$. This is a bright example of the non-commutativity of the large mass limit with the light-cone formulation of the nucleon structure operators. The Mellin moment of the expression (\ref{B5}) reads:
\begin{eqnarray}
B_N&=&\int_{-1}^1 dx~x^N~B(x)=\frac{\bar N\tau^a\hat n N}{P_+} ~\frac{g_a}{2}\frac{m^2}{(4\pi F_\pi)^2}~\chil{\Delta q}_N \tilde B_N,
\end{eqnarray}
where
$$\chil{\Delta q}_N=\int_{-1}^1d\beta~\beta^N~ \chil{\Delta q}(\beta),$$
$$
\tilde B_N=\int_0^1 dy~y^N\(\frac{1}{2}\delta(1-y)\ln(m^2)+ \frac{\bar y}{\alpha^2}
 \ln\(1+\frac{y\alpha^2}{\bar y^2}\)\).
$$
The leading term of the expansion for $\tilde B_N$ in powers of $\alpha$ vanishes due to cancelation of integrals over logarithm and delta-function. Therefore, the non-analytical part Mellin moment of the mixing parity term is
\begin{eqnarray}
B_N=\mathcal{O}(\alpha^4).
\end{eqnarray}
This result coincides with the results of \cite{Diehl:2006js,Diehl:2006ya,Ando:2006sk,Dorati:2007bk,Diehl:2005rn}. The analytical part of the diagram is of order $\mathcal{O}(\alpha^2)$ and accompanied by the leading contribution of higher chiral order operator. The mixing of vector and axial PDFs, which is  of order $\alpha^4\simeq 5\cdot10^{-4}$ as follows from the Mellin moment calculation, is of order $\alpha^2\simeq 2\cdot10^{-2}$. In the isoscalar case, where this diagram is enforced by the chiral algebra factor, the correction is of order $10\%$ for $x\sim 0.8$.

\section{Leading chiral correction to nucleon PDF}
We have  calculated  the leading order non-analytical corrections to the nucleon
parton distributions  $q^I(x,\Delta^2)$, $\Delta q^I(x,\Delta^2)$
and $E^I(x,\Delta^2)$. The results were obtained  in the Lorentz
invariant theory and then reduced to the heavy baryon limit, as it
is described in the previous section. Here we present the full set of results depending on
the isospin and on the Lorentz structure. We use the notation $\mathcal{O}(\text{NLog})$ for the reminder of the expansion. It indicates that the calculated by us terms corresponds to the leading logarithmical contribution for the Mellin moments, although they contains the non-logarithmical functions. The finite part of expression contains the tree expression for the higher derivative GPD operators. They are not presented here. For the interested reader we present
diagram-by-diagram listing in the appendix A.

In order to present the result in more convenient way, we use the
following notations: for the vector PDFs
\begin{eqnarray}\label{q0}
q^I(x,\Delta^2)&=&\chilI{q}(x)+\frac{1}{(4\pi F_\pi)^2}\int_{-1}^1 \frac{d\beta}{\beta} \theta\(0<\frac{x}{\beta}<1\)
\\&&\nn\times
 \(\chilI{q}(\beta) C^I\(\frac{x}{\beta},\Delta^2\)+\chilI{\Delta q}(\beta)\Delta C^I\(\frac{x}{\beta}\)
+\chilI{Q}(\beta)C^I_\pi\(\frac{x}{\beta},\Delta^2\)\),~~~~~~~~~~~~
\end{eqnarray}
\begin{eqnarray}\label{q0.1}
 E^I(x,\Delta^2)=\frac{1}{(4\pi
F_\pi)^2}\int_{-1}^1 \frac{d\beta}{\beta}
\theta\(0<\frac{x}{\beta}<1\)
 \(\chilI{q}(\beta) S^I\(\frac{x}{\beta},\Delta^2\)+\chilI Q(\beta) S^I_\pi\(\frac{x}{\beta},\Delta^2\)\),
\end{eqnarray}

and for the axial PDF

\begin{eqnarray}
\Delta q^I(x,\Delta^2)&=&\chilI{\Delta q}(x)
\\&&\nn+
\frac{1}{(4\pi F_\pi)^2}\int_{-1}^1 \frac{d\beta}{\beta}
\theta\(0<\frac{x}{\beta}<1\) \(\chilI{\Delta q}(\beta) \widetilde
C^I\(\frac{x}{\beta},\Delta^2\)+\chilI{q}(\beta)\Delta \widetilde
C^I\(\frac{x}{\beta}\)\),
\end{eqnarray}
where $\chilI q(x)$ and $\chilI{\Delta q}(x)$ are the nucleon PDFs
in the chiral limit, and $\chilI Q(x)$ is the pion PDF in the chiral
limit. The subscript $I$ stands for the isospin. The coefficient
functions $C^I$, $\widetilde{C}^I$ and $S^I$ are listed below.

For the vector PDF the functions $C^I$ are:
\begin{eqnarray}\label{C0}
C^0(y,\Delta^2)=-\frac{3g_a^2}{2}\delta(1-y)m^2\ln\frac{m^2}{\mu^2}-3g_a^2\int_0^{\bar
y} d\eta\frac{ym^2-\eta \bar y\Delta^2}{\bar
y^2+\alpha^2y-\frac{\Delta^2}{M^2}\eta(\bar\eta-y)}+\mathcal{O}(\text{NLog}),
\\
C^1(y,\Delta^2)=-\(1+\frac{5g_a^2}{2}\)\delta(1-y)m^2\ln\frac{m^2}{\mu^2}
+g_a^2\int_0^{\bar
y} d\eta\frac{ym^2-\eta \bar y\Delta^2}{\bar
y^2+\alpha^2y-\frac{\Delta^2}{M^2}\eta(\bar\eta-y)}~~~~~~~~
\\ \nn +\mathcal{O}(\text{NLog}).
\end{eqnarray}
For the axial PDF the functions $\widetilde{C}^I$ are
\begin{eqnarray}
\widetilde C^0(y,\Delta^2)&=&C^0(y,\Delta^2)+6g_a^2M^2\bar
y\ln\(1+\frac{\alpha^2y}{\bar y^2}\)
\\ \nn &&~~~~~~~~~~~~~~~~
-6g_a^2\int_0^{\bar
y} d\eta\frac{\eta^2 \Delta^2}{\bar
y^2+\alpha^2y-\frac{\Delta^2}{M^2}\eta(\bar\eta-y)}
+\mathcal{O}(\text{NLog}),
\\ \label{TildaC1}
\widetilde C^1(y,\Delta^2)&=&C^1(y,\Delta^2)-2g_a^2M^2\bar
y\ln\(1+\frac{\alpha^2y}{\bar y^2}\)
\\ \nn &&~~~~~~~~~~~~~~~~
+2g_a^2\int_0^{\bar
y} d\eta\frac{\eta^2 \Delta^2}{\bar
y^2+\alpha^2y-\frac{\Delta^2}{M^2}\eta(\bar\eta-y)}
+\mathcal{O}(\text{NLog}).
\end{eqnarray}
The parity-mixing functions $\Delta C^I$ are the same for the axial
and vector case:
\begin{eqnarray}
\Delta C^0(y)&=&\Delta \widetilde
C^0(y)=-\frac{3g_a}{2}\delta(1-y)m^2\ln\frac{m^2}{\mu^2}-3g_aM^2\bar
y\ln\(1+\frac{y\alpha^2}{\bar y^2}\)+\mathcal{O}(\text{NLog}),
\\\label{dC1}
\Delta C^1(y)&=&\Delta \widetilde
C^1(y)=\frac{g_a}{2}\delta(1-y)m^2\ln\frac{m^2}{\mu^2}+g_aM^2\bar
y\ln\(1+\frac{y\alpha^2}{\bar y^2}\)+\mathcal{O}(\text{NLog}).
\end{eqnarray}
The pure pion contribution to the vector PDFs is given by:
\begin{eqnarray}\label{qend-1}
C_\pi^0(y,\Delta^2)&=& \frac{3g_a^2}{2}\delta(y)\int_0^1 d\eta
R(\eta)\ln\frac{R(\eta)}{\mu^2}
\\&&\nn+
6g_a^2M^2 y \ln\(1+\frac{\alpha^2\bar y}{y^2}\)+ 6g_a^2y\int_0^{\bar
y} d \eta\frac{m^2-\eta \Delta^2}{
y^2+\alpha^2\bar y-\frac{\Delta^2}{M^2}\eta(\bar\eta-y)}+\mathcal{O}(\text{NLog}),
\\\label{qend}
C_\pi^1(y,\Delta^2)&=& \(1-g_a^2\)\delta(y)\int_0^1 d\eta
R(\eta)\ln\frac{R(\eta)}{\mu^2}
\\&&\nn-
4 g_a^2M^2y \ln\(1+\frac{\alpha^2\bar y}{y^2}\)- 4g_a^2y\int_0^{\bar
y} d \eta\frac{m^2-\eta \Delta^2}{
y^2+\alpha^2\bar y-\frac{\Delta^2}{M^2}\eta(\bar\eta-y)}+\mathcal{O}(\text{NLog}).
\end{eqnarray}
The coefficient functions $S^I$ defining the vector PDF
$E(x,\Delta^2)$ are following
\begin{eqnarray}\label{Sn}
S^I(y,\Delta^2)&=&2g_a^2T^I\int_0^{\bar y}
d\eta\frac{y m^2-\Delta^2\eta(\bar \eta-y)}{\bar
y^2+\alpha^2y-\frac{\Delta^2}{M^2}\eta(\bar\eta-y)}+\mathcal{O}(\text{NLog}),\\
\label{Sp}
 S^I_\pi(y,\Delta^2)&=&4g_a^2G^I\int_0^{\bar y} d
\eta\frac{R(\eta)-y(m^2-\Delta^2\eta)}{
y^2+\alpha^2\bar y-\frac{\Delta^2}{M^2}\eta(\bar\eta-y)} +\mathcal{O}(\text{NLog}),
\end{eqnarray}
where
$$T^0=3,~~~~~T^1=-1,~~~~~G^0=3,~~~~~G^1=-2,$$
$$R(\eta)=m^2-\eta\bar \eta \Delta^2,~~~~~\alpha=\frac{m}{M}.$$

The presented results can be compared with existent in literature results for Mellin moments.  The moments $N=0$ of parton distributions
$q(x,\Delta^2)$ and $\Delta q(x,\Delta^2)$ diagram-by-diagram reproduce  the vector and axial form factors given in \cite{Bernard:1998gv}. Also,
our results reproduce the Mellin moments of $q(x,\Delta^2)$ and $\Delta q(x,\Delta^2)$ presented in \cite{Diehl:2005rn} and corresponded nucleon
gravitational form factors given in \cite{Belitsky:2002jp}. The expressions for the distribution $E(x,\Delta^2)$ have some disagreement with the
results of \cite{Diehl:2005rn,Bernard:1998gv}. The disagreement arises from the nonlocal operators with one derivative, which should be taken
into account for the distribution $E(x,\Delta^2)$, but are not considered in this article.

The last two terms in (\ref{qend-1}-\ref{qend}) have a special
behavior in $x$. At $x\sim 1$ these terms contribute to the parton
distribution $q(x,\Delta^2)$ as  $\alpha^2$, but at $x\sim \alpha$
their contribution increases and becomes of order $\alpha$. In the
region $x\sim \alpha^2$ these terms give contributions of order
$\alpha^2\ln\alpha$. The fact that corrections change their chiral
order with $x$ is a character feature of the pure pion sector.
Similar terms in nucleon-pion sector (\ref{C0}-\ref{dC1}) do not change the order of
correction with the changing of $x$.

Also we are drawing attention to the $\delta$-function contributions
in expressions (\ref{qend-1},\ref{qend}). These contributions have
the special nature, because at $x\sim \alpha^2$  they are formally
of the same order as the tree term. Together with similar
contributions from the higher orders of chiral expansion they form
the leading chiral correction to the nucleon PDF in the region
$x\lesssim \alpha^2$. The operator evidence of the these terms was
presented in sec.2.2. The resummation of the $\delta$-function
contributions in all orders of ChPT is an important task. It leads
to the model-independent picture of the nucleon at large impact
parameters.

\section{Summary}

Using nucleon-pion  ChPT we have calculated the leading non-analytic chiral corrections for nucleon PDFs with non-zero momentum transfer in
$x$-space. We have restricted ourselves to the case $\Delta_+=0$ and to the GPD operators of the leading order. We have derived the general
rules to deal with light-cone nonlocal operators for the theories with heavy particles. Our method is the generalization of EOMS scheme  (the
extended-on-mass-shell subtraction scheme by \cite{Schindler:2003xv,Fuchs:2003qc}) and it can be applied for operators with various quantum
numbers. Application of another methods for solving the hierarchy problem, such as HBChPT and infrared regularization scheme
\cite{Becher:1999he}, result to artificial singularities in the physical domain of $x$.

The Mellin moments of our calculations coincides with the known results. However, our analysis shows that the Mellin moments calculated within
standard HBChPT approach can be used for reconstruction of the original parton distribution only in the region $x\sim 1$, and fails to restore
the distribution already at $x\sim \frac{m}{M}\simeq 0.15$. Our results can be used for investigation of the distributions down to $x\sim
\frac{m^2}{M^2}\simeq 2\cdot10^{-2}$. We found that the derived corrections give the visible contribution of order $10^{-1}$.

We have shown that there exist a parity-mixing term, which mixes vector and axial PDFs. The mixing contribution is of order $a_\chi^2$, which can not be deduced from consideration of the Mellin moments. Because of uncertainties in the inverse Mellin transformation the mixing terms were believed to be of order $a_\chi^4$, which two orders of magnitude less.

The pion contributions to the vector GPD contain the term proportional to $\delta(x)$, which is a bright sign of necessity to sum up such terms in all orders of the chiral expansion. They must be resummed at $x\sim \alpha^2$, since all $\delta$-terms  are of the same order as the leading term. The resummation can be done in the leading logarithmical order with the help of the recursive equations \cite{Kivel:2008mf}. The result gives the leading asymptotic term in the long distance behavior of nucleon GPDs, see e.g.\cite{Perevalova:2011qi}.

The slope of PDFs is related to the average impact parameter squared as
\begin{eqnarray}\label{IP}
\langle b^2_\perp(x)\rangle&=&4\frac{d
q(x,\Delta^2)}{d\Delta^2}\bigg|_{\Delta^2=0}.
\end{eqnarray}
We have found, that the contribution of the pion sector of PDFs given by (\ref{qend-1}-\ref{qend},\ref{Sp}) is larger in magnitude then the contribution of the nucleon sector (\ref{C0}-\ref{TildaC1},\ref{Sn}) for $x\lesssim\alpha$, which is in a good agreement with general chiral picture of the nucleon \cite{Strikman:2009bd}. For example, average squared impact parameter $\langle b^2_\perp\rangle$  for the vector isovector parton distribution
$q^{I=1}(x,\Delta^2)$ is:
\begin{eqnarray}\label{IPnumber}
\langle b^2_\perp\rangle&=&\int_{-1}^{1}d x\langle
b^2_\perp(x)\rangle
=-0.056\mathrm{fm}^2+0.238\mathrm{fm}^2=0.182\mathrm{fm}^2,
\end{eqnarray}
where the first term comes from nucleon contribution $C^1(x,\Delta^2)$, whereas the second term comes  from the pion contribution
$C^1_{\pi}(x,\Delta^2)$.

The important in phenomenology case of non-zero $\Delta_+$, as well as complete calculation of GPDs $E$ and $\widetilde E$, also can be considered in the presented framework and will be covered in further publication.

We would like to thank M.Polyakov, N.Kivel and G.Gegelia for numerous stimulating discussions and real interest to our work. Also we thank
P.Wein for comments. This work is supported by BMBF grant 06BO9012.

\appendix
\section{List of diagram contributions}
In this appendix we present the listing of the diagram
contributions, in $x$-representation and their Mellin moments at the
leading logarithmical order.
 The diagrams are shown in fig.1. To obtain the results (\ref{q0}-\ref{qend}) the renormalization of the nucleon field should be added.

Diagrams $A,B,C,D,E$ corresponding to the matrix element
(\ref{V_QCD})  can be  presented as the following convolution:
\begin{eqnarray}
\text{Diagram}=\frac{P_+^{-1}}{(4\pi F_\pi)^2}\int_{-1}^1
\frac{d\beta}{\beta} \theta\(0\leq \frac{x}{\beta}\leq1\)\chilI
q(\beta)\bar u(p')\[\DiracSlash n
\Theta^I_H\(\frac{x}{\beta}\)+\frac{i\sigma^{\mu\nu}n_\mu\Delta_\nu}{2M}\Theta^I_E\(\frac{x}{\beta}\)\]u(p),
\end{eqnarray}
where  $\Theta^I_H(x)$ and $\Theta^I_E(x)$ belong to the nucleon  parton distributions $q^I(x,\Delta^2)$
and $E^I(x,\Delta^2)$, respectively. The vector PDF $\chilI q(\beta)$ should be replaced by
the axial PDF $\chilI{\Delta q}(\beta)$ for the diagram $B$, and by the  pion PDF $\chilI{Q}(\beta)$ for the diagrams $D$ and $E$.

Functions $\Theta^I_H$ for different diagrams read:
\begin{eqnarray*}
\Theta^I_{H,A}(y)&=&g_a^2T^I\(\delta(1-y)\frac{m^2}{4}\ln\frac{m^2}{\mu^2}-\int_0^{\bar
y} d\eta\frac{ym^2-\eta \bar y\Delta^2}{\bar
y^2+\alpha^2y-\frac{\Delta^2}{M^2}\eta(\bar\eta-y)}\)+\mathcal{O}(\text{NLog}),
\\
\Theta^I_{H,B}(y)&=&-\frac{g_aT^I}{2}\(\delta(1-y)\frac{m^2}{2}\ln\frac{m^2}{\mu^2}+M^2\bar
y\ln\(1+\frac{\alpha^2y}{\bar y^2}\)\) +\mathcal{O}(\text{NLog}),
\\
\Theta^I_{H,C}(y)&=&-\delta^{I,1}\delta(1-y)m^2\ln\frac{m^2}{\mu^2}+\mathcal{O}(\text{NLog}),\\
\Theta^I_{H,D}(y)&=&g_a^2G^I\( \frac{\delta(y)}{2}\int_0^1 d\eta
R(\eta)\ln\frac{R(\eta)}{\mu^2}+2M^2 y\ln\(1+\frac{\alpha^2\bar
y}{y^2}\)\right.
\\\nn&&~~~~~~~~~~~~~~~~~~~~~~~~~~~~~~~~~~~~~~~~~+
\left.\int_0^{\bar y} d
\eta\frac{2y(m^2-\Delta^2\eta)}{
y^2+\alpha^2\bar y-\frac{\Delta^2}{M^2}\eta(\bar\eta-y)}\)
+\mathcal{O}(\text{NLog}),\\
\Theta^I_{H,E}(y)&=&\delta^{I,1}\delta(y)\int_0^1 d\eta
~R(\eta)\ln\frac{R(\eta)}{\mu^2}+\mathcal{O}(\text{NLog}),
\end{eqnarray*}
where
$$
R(\eta)=m^2-\eta\bar \eta \Delta^2,~~~~~\alpha=\frac{m}{M},
$$
$$
T^0=3,~~~~
T^1=-1,~~~~
G^0=3,~~~~
G^1=-2.
$$

Only triangle diagrams $A$ and $D$ contribute to the distribution
$E(x,\Delta^2)$:
\begin{eqnarray*}
\Theta^I_{E,A}(y)&=&2g_a^2T^I\int_0^{\bar y}
d\eta\frac{ym^2-\Delta^2\eta(\bar \eta-y)}{\bar
y^2+\alpha^2y-\frac{\Delta^2}{M^2}\eta(\bar\eta-y)}+\mathcal{O}(\text{NLog}),
\\
\Theta^I_{E,D}(y)&=&4g_a^2G^I\int_0^{\bar y} d
\eta\frac{R(\eta)-y(m^2-\Delta^2\eta)}{
y^2+\alpha^2\bar y-\frac{\Delta^2}{M^2}\eta(\bar\eta-y)} +\mathcal{O}(\text{NLog}).
\end{eqnarray*}

Three diagrams  $A,B,C$ contributing to the axial matrix element
(\ref{A_QCD}) will be presented in a similar convolution:
\begin{eqnarray}
\text{Diagram}=\frac{\bar u(p')\DiracSlash n\gamma_5u(p)}{(4\pi
F_\pi)^2P_+}\int_{-1}^1 \frac{d\beta}{\beta}
\theta\(0\leq\frac{x}{\beta}\leq1\)\chilI {\Delta q}(\beta)
\widetilde{\Theta}^I_H\(\frac{x}{\beta}\),
\end{eqnarray}
where  $\widetilde{\Theta}^I_H(x)$ belongs to  the
nucleon  parton distribution $\Delta q^I(x,\Delta^2)$. The axial PDF
$\chilI {\Delta q}(\beta)$ should be replaced by the vector PDF $\chilI q(\beta)$ for the diagram $B$.
\begin{eqnarray*}
\widetilde{\Theta}^I_{H,A}(y)&=&g_a^2
T^I\(\delta(1-y)\frac{m^2}{4}\ln\frac{m^2}{\mu^2}+2M^2\bar y\ln\(1+\frac{\alpha^2 y}{\bar
y^2}\)\)
\\
&&\nn
~~~~~~~~~~~~~~~~~~~~~~~~-\int_0^{\bar y}d\eta \frac{ym^2+(4\eta^2-\bar y^2) \frac{\Delta^2}{2}}{\bar y^2+\alpha^2 y-\frac{\Delta^2}{M^2} \eta (\bar y-\eta)}
+ \mathcal{O}(\text{NLog}),
\\
\widetilde{\Theta}^I_{H,B}(y)&=&-\frac{g_aT^I}{2}\(\delta(1-y)\frac{m^2}{2}\ln\frac{m^2}{\mu^2}+M^2\bar
y\ln\(1+\frac{\alpha^2y}{\bar y^2}\)\) +\mathcal{O}(\text{NLog}),
\\
\widetilde{\Theta}^I_{H,C}(y)&=&-\delta^{I,1}\delta(1-y)m^2\ln\frac{m^2}{\mu^2}+\mathcal{O}(\text{NLog}).
\end{eqnarray*}

Below we present non-analytic contributions to  Mellin moments of $q(x,\Delta^2)$, $\Delta q(x,\Delta^2)$ and $E(x,\Delta^2)$ nucleon parton distributions at order $\mathcal{O}(\alpha^2)$ in the chiral expansion.  We define Mellin transformation of the functions $\Theta^I_{H}(y)$, $\Theta^I_{E}(y)$ and $\widetilde{\Theta}^I_{H}(y)$ as
\begin{align}
&\Theta^I_{H}(N)=\int_0^1dy\, y^N\Theta^I_{H}(y),&&
\Theta^I_{E}(N)=\int_0^1dy\,y^N\Theta^I_{E}(y),&&
\widetilde{\Theta}^I_{H}(N)=\int_0^1dy\,
y^N\widetilde{\Theta}^I_{H}(y),
\end{align}
 and give a list of
their moments diagram-by-diagram:
\begin{eqnarray*}
\Theta^I_{H,A}(N)&=&\frac{3g_a^2}{4}T^Im^2\ln\frac{m^2}{\mu^2}+\mathcal{O}(\alpha^4),
\\
\Theta^I_{H,B}(N)&=&\mathcal{O}(\alpha^4),
\\
\Theta^I_{H,C}(N)&=&-\delta^{I,1}m^2\ln\frac{m^2}{\mu^2}
+\mathcal{O}(\alpha^4),
\\
\Theta^I_{H,D}(N)&=&\delta^{I,1}\delta_{N,0}g_a^2\int\limits_0^1d\eta\(\(m^2-\Delta^2\eta(1+\eta)\)\ln\frac{R(\eta)}{\mu^2}+2m^2\ln\frac{m^2}{\mu^2}\)
 +\mathcal{O}(\alpha^4),
\\
\Theta^I_{H,E}(N)&=&\delta^{I,1}\delta_{N,0}\int_0^1 d\eta
~R(\eta)\ln\frac{R(\eta)}{\mu^2}+\mathcal{O}(\alpha^4),
\end{eqnarray*}
\begin{eqnarray*}
\widetilde{\Theta}^I_{H,A}(N)&=&-\frac{g_a^2}{4}
T^Im^2\ln\frac{m^2}{\mu^2}+\mathcal{O}(\alpha^4),
\\
\widetilde{\Theta}^I_{H,B}(N)&=&\mathcal{O}(\alpha^4),
\\
\widetilde{\Theta}^I_{H,C}(N)&=&-
\delta^{I,1}m^2\ln\frac{m^2}{\mu^2}
+\mathcal{O}(\alpha^4).~~~~~~~~~~~~~~~~~~~~~~~~~~~~~~~~~~~~~~~~~~~~
\end{eqnarray*}
\begin{eqnarray*}
\Theta^I_{E,A}(N)&=&-g_a^2T^Im^2\ln\frac{m^2}{\mu^2}+\mathcal{O}(\alpha^4),
\\
\Theta^I_{E,D}(N)&=&-4g_a^2\delta^{I,1}\delta_{N,0}\int\limits_0^1d\eta\(\pi
M\sqrt{R(\eta)}+\(m^2-\Delta^2\eta\)\ln\frac{R(\eta)}{\mu^2}\)\nn\\
&&-6g_a^2\delta^{I,0}\delta_{N,1}\int\limits_0^1d\eta
R(\eta)\ln\frac{R(\eta)}{\mu^2}+\mathcal{O}(\alpha^4),~~~~~~~~~~~~~~~~~~~~~~~~~~~~~~~~~~
\end{eqnarray*}
Functions $\Theta^I_H(N=0)$ and $\widetilde{\Theta}^I_H(N=0)$ for
zeroth Mellin moment diagram-by-diagram reproduce results of
\cite{Bernard:1998gv} for the vector and axial form factors.
Functions $\Theta^I_E(N=0)$ differ in the logarithm parts.

\end{document}